\documentclass{article}
\usepackage[utf8]{inputenc}
\usepackage{geometry}
\usepackage{physics}
\usepackage{dsfont}
\usepackage{amsmath}
\usepackage{amssymb}
\usepackage{graphicx}
\usepackage{babel,blindtext}
\usepackage{xcolor}
\usepackage{hyperref}
\usepackage{cite}
\usepackage{wrapfig}
\usepackage{caption}
\usepackage{subcaption}
\usepackage{bbm}
\usepackage{authblk}
\title{Affinity-based geometric discord and quantum speed limits of its creation and decay}

\date{}
\author[1]{\small Rajendran Muthuganesan\footnote{Corresponding author}}
\author[2]{\small S. Balakrishnan}
\affil[1]{\footnotesize Department of Physics, Faculty of Nuclear Sciences and Physical Engineering, Czech Technical University in Prague,
B\u rehov\'a 7, 115 19 Praha 1-Star\'e M\u{e}sto, Czech Republic, Email: rajendramuthu@gmail.com}
\affil[2]{\footnotesize Department of Physics, School of Advanced Sciences, Vellore Institute of
Technology, Vellore– 632014, India.}
\begin{document}

\maketitle
%%%%%%%%%%%%%%%%%%%%%%%%%%%%%%%%%%%%%%%%%%%%%%%%%%%%%%%%%%%%%%
\begin{abstract}
In this article, we define a faithful quantifiers of bipartite quantum correlation, namely geometric version of quantum discord using affinity based metric. It is shown that the newly-minted measure resolves the local ancilla problem of Hilbert-Schmidt measures. Exploiting the notion of affinity-based discord, we derive  Margolus-Levitin (ML) and Mandelstamm-Tamm (MT) bounds for the quantum speed limit time for the creation and decay of quantum correlation. The dynamical study suggests that the affinity measure is a better resource compared to entanglement. Finally, we study the role of quantum correlation on quantum speed limit.  
\vspace{0.75cm}

\noindent Keywords : Quantum speed limit; Quantum correlationa; Affinity; Uncertainty relation; Ornstein-Uhlenbeck noise.
\end{abstract}

\section{Introduction}
\par Entanglement and uncertainty, foundational aspects of the physical systems in the quantum regime, are the building blocks for the development of modern quantum technology. Entanglement is a peculiar phenomena emerges from the superposition of quantum states and found to be  an  important for advantages of various quantum tasks such as teleportation, super dense coding, quantum communication etc \cite{Nielsen2010}. Owing to the rigorous framework of quantum resource theory,  different entanglement measures have been proposed in both bipartite and multipartite scenarios \cite{Plenio2007}.  Over the past decades, the research have been carried out on the nonlocal features and the analysis centered around the violation of Bell inequality \cite{Schrodinger1935,Schrodinger1936,Bell}. The characterization of quantum systems from the perspective of nonlocality is a formidable task and  reveals that the entanglement is  not a whole indicators of the nolocal attributes of the quantum system \cite{Werner1989}. Further development in the quantum resource theory identifies the different measures such as quantum discord \cite{Ollivier2001}, measurement-induced nonlocality \cite{Luo2011}, quantum coherence \cite{Baumgratz2014,Girolami2014} etc.

On the other hand, the uncertainty relation (UR), originally introduced by Heisenberg, places the accuracy while measuring the observable \cite{Auletta}. In other words, it is shown that the outcomes for two incompatible measurements cannot be accurately measured at the same time. In recent times, the entropy based uncertainty relation is powerful tool in the characterization of quantum system dynamics  in the different environment. Later, energy-time UR also interpreted as quantum speed limits (QSLs) of unitary evolution of physical systems and  QSL is defined as the bounds on the shortest time required to evolve between two quantum states \cite{MT}.

QSL provides a key understanding on the evolution and controlling of quantum system,  and also unifies the framework of the both computational and evolution speed. Due to the wide range of applications of QSL in quantum information theory, the researchers got enormous interest on characterization of QSL for nonunitary dynamics.  QSL has been studied from the different perspectives such as  decay rate of quantum states, parameter estimation, rate of information transfer and processing, entropy production [28], precision in quantum metrology, state preparation  and time scales of quantum optimal control, many-body open system \cite{Bekenstein1981,Lloyd,Giovannetti2011,Deffner2010,Caneva2009,Fleming1973,Bhattacharyya1983,Anandan1990,
Vaidman1992,Uffink1993,Brody2003,Luo2005LMP,Boixo2007,Aifer,Bukov}. Recently, QSL for the states with a bounded energy spectrum is also studied \cite{Nes} and realizable using different candidates such as on single photons \cite{Langford}, trapped ions \cite{Klimov}, trapped atoms \cite{Belmechri}, and artificial atoms in superconducting quantum circuits \cite{XYX,Cervera}.  In recent times, to quantify the maximal quantum speed of systems interacting with their environments different approaches have been proposed in terms of Fisher information \cite{Taddei2013}, relative purity \cite{Campo2013} and geometric view point \cite{Deffner2013}. It is shown that the  entanglement “speed up” the time evolution of composite systems \cite{Giovannetti2003a,Giovannetti2003b,Zander2007}.  Identifying the relation between QSL and resource theory is a promising field with in the framework of QIT.  Ornstein-Uhlenbeck (OU) noise is an important non-Markovian decoherence in the framework of quantum information theory which provides a mathematical description of Brownian motion \cite{Uhlenbeck}.   The Ornstein–Uhlenbeck noise has been widely studied in different perspectives \cite{Chiara2003,Fiasconaro2009,Grotz2006,Altintas2020}. The OU noise could be helpful to control the loss of quantum resources in quantum system's dynamical maps.  Hence, the understanding of the loss of information and control of resources under OU noise  is very essential in the context of quantum information theory.

In view of this, in this present work, we derive Margolus-Levitin (ML) and Mandelstamm-Tamm (MT) like bound of QSL time for the decay and creation of quantum correlation using newly minted affinity based geometric discord. The affinity based quantum discord measure resolves the local ancilla problem of Hilbert-Schmidt based quantifiers \cite{Luo2011,geometric1}. It is shown that the robustness of affinity-based discord under Ornstein-Uhlenbeck (OU) dephasing noise suggest it would be a better resource than the entanglement. We investigate the QSL time for the decay and creation of quantum correlations for the nonunitary evolution under OU dephasing noise.  

 The  paper is structured  as follows. In Sec. \ref{sec2}, first we review the notion of quantum discord, then we define affinity-based geometric version of quantum discord.  In Sec.  \ref{sec3}, we derive ML and MT like bound of QSL time through the newly-minted discord. The dynamics of quantum correlation under OU dephasing noise and QSL time are calculated in Sec.  \ref{sec4}. Finally, the conclusions are presented in Sec.  \ref{cncl}.

\section{Affinity-based discord}
\label{sec2}
It is widely accepted fact that the entanglement could not capture all the nonlocal aspects of a quantum system. To provide complete the spectrum of the nonlocal aspects of quantum system, Olivier and Zurek identified a new quantifiers from the perspective measurement is called quantum discord (QD). First, we recall some basic notations and definitions related the discord measure.  Let $\rho_{ab}$ be a two-qubit state shared between the parties $a$ and $b$ in the separable Hilbert space $\mathcal{H}^a\otimes\mathcal{H}^b$.  Then the mutual information of $\rho_{ab}$ is defined as 
\begin{align}
\mathcal{I}(\rho_{ab})=S(\rho^a)+S(\rho^b)-S(\rho_{ab}) 
\label{mutual}
\end{align}
where $\rho^{i}=\text{Tr}_j(\rho_{ab})~ (\{i,j \} \in \{a,b \}, i\neq j)$ are the marginal states of $\rho_{ab}$. Here, $S(\rho)=-\text{Tr}\rho\text{log}\rho$ is von Neumann entropy of quantum state and logarithm is base 2. The quantum conditional entropy of the bipartite state is given by 
\begin{align}
  S_{a|b}=~^{\text{min}}_{\Pi ^{b}_k \in \mathcal{M}}\sum_k p_kS(\rho_{a|k})  \nonumber
\end{align}
where the minimization is taken over all quantum measurements $\{ \Pi ^{b}_k\} $ on $b$ and $\mathcal{M}$ is set of all such measurements. The post-measurement state is $\rho_{a|k}=\text{Tr}_b[(\mathds{1}\otimes\Pi_k^{b})\rho_{ab}(\mathds{1}\otimes\Pi_k^{b})]/p_k$ with $p_k=\text{Tr}(\mathds{1}\otimes\Pi_k^{b})\rho_{ab}(\mathds{1}\otimes\Pi_k^{b})$. The equivalent expression of mutual information eq. (\ref{mutual}), defined as 
\begin{align}
  \mathcal{J}_{a|b}=S(\rho^a)-S_{a|b}.  
  \label{mutual2}
\end{align}
The mutual information eq. (\ref{mutual}) is always greater than eq. (\ref{mutual2}) and both are equal in the classical regime. The QD is defined as \cite{Ollivier2001}
\begin{align}
  \mathcal{D}=\mathcal{I}(\rho_{ab})-\mathcal{J}_{a|b}
\end{align}
the difference between two inequivalent expressions of mutual information in the quantum regime. The above quantity also demonstrate the presence of correlation in separable state. Even though, the complete manifestation  of nonlocality, the computation of QD is a non-deterministic polynomial even for a simple bipartite state.  

 Motivated by the notion of quantum discord, the geometric version of discord is defined as \cite{geometric1,geometric2}
\begin{align}
   \mathcal{D}_G(\rho_{ab})=~~^{\text{min}}_{\{ \Pi_k^{a}\}} \lVert \rho_{ab}-\Pi^{a}(\rho_{ab})\rVert^2  \label{GD2}  
\end{align}
where the optimization is taken over von Neumann projective measurements and $\Pi^{a}(\rho)= \sum_k(\Pi_k^{a}\otimes \mathds{1})\rho (\Pi_k^{a}\otimes \mathds{1})$ with $\{ \Pi_k^{a}\} =\{ |k\rangle \langle k|\} $ are the projectors acting on state space $a$ and $\mathds{1}$ is the identity operator in state space $b$. Here $\lVert \mathcal{O}\rVert^2=\text{Tr}\mathcal{O}\mathcal{O}^{\dagger}$ is the Hilbert-Schmidt norm of operator $\mathcal{O}$. The quantum discord based on the Hilbert-Schmidt is not a faithful measure of quantum correlation due to the non-cotractvity and local ancilla problem. To combat this local ancilla issue, the different versions of geometric discord is studied using Schatten 1-norm \cite{Paula2013}, Bures distance \cite{Spehner2013} and Hellinger distance \cite{Roga}. 

In this context, we introduce a new version of geometric QD based on affinity as
\begin{align}
  \mathcal{D}(\rho_{ab})=~~^{\text{min}}_{\{\Pi_k^{a}\}} d_{\mathcal{A}}(\rho_{ab}, \Pi^{a}(\rho_{ab}))
  \label{affinitygd}
\end{align}
where $\mathcal{A}(\rho, \sigma)=\text{Tr}\sqrt{\rho}\sqrt{\sigma}$ is the affinity between the states $\rho$ and $\sigma$ \cite{Holevo1972,Affinitydef} and $d_{\mathcal{A}}(\rho, \sigma)=1-\sqrt{\mathcal{A}(\rho, \sigma)}$ is a affinity-induced metric. In recent times, the usefulness of affinity in characterization of quantum correlation of bipartite state \cite{GDaff}, bilocal state \cite{bilocalaff} and quantum coherence  \cite{coh1aff,coh2aff} is also demonstrated. A similar quantum correlation measure is characterized using Hellinger distance \cite{Roga}.  The properties of the  affinity-based measure are

\begin{enumerate}
\item[(i)]  $\mathcal{D}(\rho_{ab})$ is non-negative and positive for entangled state. 

\item[(ii)] $\mathcal{D}(\rho_{ab})=0$ for any product state  and the classical-quantum state in the form $\rho_{ab}=\sum _{k}p_{k}|k\rangle \langle k| \otimes   \rho_{k}  $ with nondegenerate marginal state $\rho^{a}=\sum_{k}p_{k}|k\rangle \langle k|$.

\item[(iii)] $\mathcal{D}(\rho_{ab})$ is locally unitary  invariant i.e., $\mathcal{D}\left((U\otimes   V)\rho_{ab}  (U\otimes   V)^\dagger\right)=\mathcal{D}(\rho_{ab})$ for any local unitary operators $U$ and $V$.

\item[(iv)] $\mathcal{D}(\rho_{ab})$ is invariant under the addition of any local ancilla to the unmeasured party $b$. 

\item[(v)] For any arbitrary pure state $|\Psi\rangle=\sum_is_i |\alpha_i\rangle \otimes  |\beta_i\rangle $ with $m\leq n$, the $\mathcal{D}(|\Psi\rangle \langle \Psi|)=1-\sqrt{\sum_is_i^2}$.

\item[(vi)]  Any $2\times n$ dimensional mixed state, the measure $\mathcal{D}(\rho_{ab})$ has a closed formula:
\begin{align}
   \mathcal{D}(\rho_{ab})=1-~~^{\text{max}}_{r_{ki}} \sqrt{r_{ki}T_{ij}r_{ki}}=1-\sqrt{T_1}
\end{align}
where $T_1$ is the latgest eigenvalue of the matrix $T$ with the elements $T_{ij}=\text{Tr}[\sqrt{\rho_{ab}} (\sigma_i\otimes \mathds{1})\sqrt{\rho_{ab}} (\sigma_j\otimes \mathds{1})] $.
\end{enumerate}

The properties (i) - (iv) guarantee that the affinity-based discord is legitimate measure of quantum correlation. 

 Now, we proceed to sketch the proof of the above properties. The property (i) is straight forward to understand. To prove the property (ii), we consider the eigenprojectors of the marginal state $\rho^a=\text{Tr}_b(\rho_{ab})=\sum_{k}p_{k}|k\rangle \langle k|$ as $\{ \Pi_k^a\} =\{ |k\rangle \langle k|\} $. Then the state and measured state for the product and classical-quantum state are equal i.e., $\Pi^a(\rho_{ab})=\rho_{ab}$ and the desired results follows. Property (iii) follows from the unitary invarience of affinity.

In order to prove the property (iv), it is sufficient to prove the affinity do not affected by the addition of local ancilla. First, we define a map $\Gamma ^\sigma: X\rightarrow X \otimes\sigma$, i.e., the map adding a noisy ancillary state to the party $b$. Under such an operation \cite{Piani2012,Chang2013}
\begin{align}
\| X \| \rightarrow \| \Gamma ^\sigma X \|=\| X \| \sqrt{\text{Tr}\sigma^2}.
\end{align}
Due to the addition of local ancilla $\rho^c$, the resultant state is $\rho_{a:bc}=\rho_{ab}\otimes \rho_c$. Here : represents the partition between the system. Hence the geometric discord eq.(\ref{GD2} ) of the resultant state is
\begin{align}
  \mathcal{D}_G(\rho^{a:bc} ) =  \mathcal{D}_G(\rho^{ab}) \cdot \text{Tr}(\rho ^{c})^2  \nonumber
\end{align}
implying that Hilbert-Schmidt based discord is not an ideal candidate in measuring quantum correlation. To show the affinity based measure resolves this issue, we recall the multiplicative property of the affinity: 
\begin{align}
\mathcal{A}(\rho_1 \otimes \rho_2,\sigma_1 \otimes \sigma_2)=\mathcal{A}(\rho_1 \otimes\sigma_1)\cdot \mathcal{A}(\rho_2 \otimes\sigma_2). \label{multiaffinity}
\end{align}
The affinity between the pre-- and post--measurement states is written as
\begin{align}
\mathcal{A}\left(\rho_{a:bc},\Pi ^{a}(\rho_{a:bc})\right) =\mathcal{A}\left(\rho_{ab}\otimes  \rho ^{c},\Pi ^{a}(\rho_{ab})\otimes  \rho ^{c}\right). \nonumber
\end{align}
Using properties of affinity such as multiplicativity and the affinity of same state is unity, we show that
\begin{align}
 \mathcal{A}\left(\rho_{a:bc},\Pi ^{a}(\rho_{a:bc})\right)=\mathcal{A}\left(\rho_{ab}, \Pi ^{a} (\rho_{ab})\right), \nonumber  
\end{align}
the affinity do not altered by the addition of local ancilla.  Here, the contractivity of the distance measure $d_{\mathcal{A}}(\rho, \sigma)$ is shown in the context of local ancilla problem. In general, we can show that contractivity of the  affinity distance from the monotonicity property of the affinity \cite{coh2aff}. Hence the  contractive nature evident that the  affinity based measure is a faithful measure of quantum correlation.

To show the property (v), we follow the Schmidt decomposition of the pure state as $|\Psi\rangle=\sum_i\sqrt{s_i} |\alpha_i\rangle \otimes  |\beta_i\rangle $ and the corresponding density matrix is $\rho=|\Psi\rangle\langle \Psi|=\sum_{ij} \sqrt{s_is_j}|\alpha_i\rangle\langle \alpha_j| \otimes |\beta_i\rangle\langle \beta_j|$. The von Neumann projective measurements on the subsystem $a$ can be represented as $\{ \Pi_k^a\} =\{ U|\alpha_k\rangle \langle \alpha_k|U^{\dagger}\} $ and the subsystem $\rho^a$ can be written as 
\begin{align}
  \rho^a=\sum_{k}\langle \alpha_k |U^{\dagger}\rho^aU| \alpha_k \rangle U|\alpha_k\rangle \langle \alpha_k|U^{\dagger}
\end{align}
the spectral decomposition in the measurement basis $\{ U|\alpha_k\rangle\} $ with the eigenvalues $\langle \alpha_k |U^{\dagger}\rho^aU| \alpha_k \rangle $. 

Exploiting the identity $\Pi^af(\rho)\Pi^a=f(\Pi^a\rho\Pi^a)$ \cite{Girolami2012}, the square root of post-measurement state is written as $\sqrt{\Pi^a(\rho_{ab})}=\Pi^a(\sqrt{\rho_{ab}})$. After the straight forward algebra, the affinity between the state under our consideration and the measured state is computed as 
\begin{align}
  \mathcal{A}(\rho_{ab},\Pi^a(\rho_{ab}))=\sum_k s_k^2.
\end{align}
Then the affinity based geometric discord for pure state is computed as
\begin{align}
  \mathcal{D}(\rho_{ab})=1-\sqrt{\sum_k s_k^2}.
\end{align}

To derive the closed formula of affinity based measure (property vi), we define the mixed quantum state in the separable Hilbert space $\mathcal{H}^{a}\otimes \mathcal{H}^{b}$. Assume $\{X_{i}:i=0,1,2,\cdots,m^{2}-1\} \in \mathcal{L}(\mathcal{H}^a)$ be a set of orthonormal operators for the state space $\mathcal{H}^a$ and $\{Y_{j}:j=0,1,2,\cdots,n^{2}-1\}  \in \mathcal{L}(\mathcal{H}^b)$ for the state space $\mathcal{H}^b$. The operators $X_{i}$ and $Y_{j}$ are satisfying the conditions $\text{Tr}(X_{k}^{\dagger }X_{l})=\text{Tr}(Y_{k}^{\dagger}Y_{l})=\delta _{kl}$. Any arbitrary $m\times n$ dimensional state of a bipartite system in the composite space  $\mathcal{L} (\mathcal{H}^{a}\otimes \mathcal{H}^{b}) $ can be written as 
\begin{equation}
\sqrt{\rho_{ab}}= \sum_{i,j}\gamma _{ij}X_{i}\otimes  Y_{j}, \label{c}
\end{equation}
where $\Gamma$ is the real matrix with the elements $\gamma _{ij} =\text{Tr} (\sqrt{\rho} ~X_{i}\otimes  Y_{j})$.  After a straight forward calculation, we compute the affinity between pre- and post- measurement state as 
\begin{align}
\mathcal{A}(\rho_{ab},\Pi^a(\rho_{ab}))=\text{Tr}(R\Gamma\Gamma^tR^t).
\end{align}
where $R$ is the measurement operators with $r_{ki}=\text{Tr}(|k \rangle \langle k|X_i)$. Then the affinity based measure is 
\begin{equation}
\mathcal{D}(\rho_{ab})=1-~~^{\text{max}}_{R} \sqrt{\text{Tr}(R\Gamma\Gamma^tR^t)}.
\label{FR}
\end{equation}
In order to obtain the closed formula for any $2\times n$ dimensional mixed state, we follow the optimization technique given in Ref.\cite{Chang2013}. The Bloch representation of the eigenprojectors is given by
\begin{align}
  \Pi_k^a=\frac{1}{2}[\mathds{1}_2+\vec{r} \cdot\vec{\sigma}  ] \nonumber
\end{align}
with $\vec{r} \cdot\vec{\sigma}=\sum_{i=1}^3 r_{ki} \sigma_i$ and $\sum_ir_{ki}=1$. Then the measurement operator $R$ is 
\begin{align}
  R=
\begin{pmatrix}
1 & r_{01} & r_{02} & r_{03} \\
1 & -r_{01} & -r_{02} & -r_{03}
\end{pmatrix}
\nonumber
\end{align}
and substituting in eq. (\ref{FR}), we obtain the closed formula of affinity based discord as 
\begin{align}
   \mathcal{D}(\rho_{ab})=1-~^{\text{max}}_{r_{ki}} \sqrt{r_{ki}T_{ij}r_{ki}}=1-\sqrt{T_1}
\end{align}
where $T_1$ is the largest eigenvalue of the matrix $T$ with the elements $T_{ij}=\text{Tr}[\sqrt{\rho_{ab}} (\sigma_i\otimes \mathds{1})\sqrt{\rho_{ab}} (\sigma_j\otimes \mathds{1})] $. Due to the analytical formula, the affinity based correlation measure is easy to compute while compared with the quantum discord.

\section{Quantum Speed Limit}
\label{sec3}
In this section, we aim to define quantum speed time $\tau$ for the decay or creation of quantum correlation in a system for a non-unitary evolution. To derive the Margolus-Levitin (ML) and Mandelstamm-Tamm (MT) bounds for the quantum speed limit time for the creation and decay of quantum correlation, we follow the procedure given in Ref. \cite{Paulson2022}. It is defined as the minimum time required to change by an amount 
\begin{align}
  \Delta\mathcal{Q}=\lvert \mathcal{D}(\rho_0) - \mathcal{D}(\rho_t) \rvert 
\end{align}
is the difference between the initial state correlation $\mathcal{D}(\rho_0)$ and final state correlation $\mathcal{D}(\rho_t)$  after the nonunitary evolution. Here, we consider the nonunitary evolution of the form 
\begin{align}
  \dot{\rho}_t=\mathcal{L}_t\rho_t   \nonumber
\end{align}
where $\mathcal{L}_t$ is the generator of the evolution. In order to quantify the  quantum speed limit (QSL) time for  $\Delta\mathcal{Q}$, we exploit the notion of affinity-based geometric discord introduced in Eq. (\ref{affinitygd}). The difference between discord of the initial and final state after the nonunitary evolution  can be written as 
\begin{equation}
\Delta\mathcal{Q} =
\begin{cases}
\sqrt{\mathcal{A}(\rho_t)}- \sqrt{\mathcal{A}(\rho_0)}~~~~
 \text{if}~~~~~~~~ \mathcal{D}(\rho_0) \geq   \mathcal{D}(\rho_{t}),\\
 \sqrt{\mathcal{A}(\rho_0)}- \sqrt{\mathcal{A}(\rho_t)}~~~~ ~~ ~~~~~~~~\text{otherwise}.
\end{cases} 
\label{HSMIN}
\end{equation}
 Here $\mathcal{A}(\cdot)$ represents the optimal affinity between the state under our consideration and  corresponding measured state. From the above, the temporal rate of change of $\Delta\mathcal{Q} $ is computed as 
\begin{align}
  \frac{d}{dt}\Delta\mathcal{Q} = \frac{\dot{\mathcal{A}}(\rho_t)}{2\sqrt{\mathcal{A}(\rho_t)}}.  \nonumber
\end{align}
It is obvious that $\frac{d}{dt}\Delta\mathcal{Q}\leq \lvert \frac{d}{dt}\Delta\mathcal{Q} \rvert $, then 
\begin{align}
2(1-\mathcal{D}(\rho_t))  \frac{d}{dt}\Delta\mathcal{Q} \leq \lvert \dot{\mathcal{A}}(\rho_t)\rvert.
\end{align}
Let $\rho_{\psi}(t)=| \psi(t)\rangle \langle \psi(t)|$ and $\sigma_{\phi}(t)=| \phi(t)\rangle \langle \phi(t)|$ are the purification of the states $\rho$ and $\Pi^a(\rho)$ respectively. Then, we have 
\begin{align}
2(1-\mathcal{D}(\rho_0)\mp\Delta\mathcal{Q})  \frac{d}{dt}\Delta\mathcal{Q} \leq \lvert \langle \phi|\mathcal{L}_t\rho_{\psi}(t)^{1/2}|\phi\rangle + \langle \psi|\mathcal{L}_t\sigma_{\phi}(t)^{1/2}|\psi\rangle \rvert
\label{Boundeq}
\end{align}
where $\mp$ indicates the decay or creation of quantum correlation under nonunitary evolution and $\mathcal{L}_t$ is generator of the evolution. 
 \subsection{Margolus-Levitin bound:}
Using the trace property $\text{Tr}|m\rangle \langle n|A=\langle n | A| m\rangle $, the above equation can be recast as 
\begin{align}
2(1-\mathcal{D}(\rho_0)\mp\Delta\mathcal{Q})  \frac{d}{dt}\Delta\mathcal{Q} \leq \lvert \text{Tr} \{ |\phi\rangle\langle \phi|\mathcal{L}_t\rho_{\psi}(t)^{1/2}\}  + \text{Tr} \{ |\psi\rangle\langle \psi|\mathcal{L}_t\sigma_{\phi}(t)^{1/2} \} \rvert
\end{align}
Let $A$ and $B$ are the two complex matrices. Then the von Neumann trace inequality is $\text{Tr}|AB|\leq \sum_{i=1}^n\lambda_{1,i}\lambda_{2,i}$ with $\lambda_{1,i}(\lambda_{2,i})$ are the singular values of $A(B)$ arranged in decreasing order and using the triangle inequality of absolute values, the above inequality can be written as 
\begin{align}
2(1-\mathcal{D}(\rho_0)\mp\Delta\mathcal{Q})  \frac{d}{dt}\Delta\mathcal{Q}\leq \lambda_{1,1}+ \lambda_{2,1}
\end{align}
where $\lambda_{1,i}(\lambda_{2,i})$ is the singular values of $\mathcal{L}_t\rho_{\psi}(t)^{1/2}(\mathcal{L}_t\sigma_{\phi}(t)^{1/2})$. Then, we have 
\begin{align}
2(1-\mathcal{D}(\rho_0)\mp\Delta\mathcal{Q})  \frac{d}{dt}\Delta\mathcal{Q}\leq \|\mathcal{L}_t\rho_{\psi}(t)^{1/2} \|_{op} +\|\mathcal{L}_t\sigma_{\phi}(t)^{1/2}) \|_{op} \\
\leq \|\mathcal{L}_t\rho_{\psi}(t)^{1/2} \|_{tr} +\|\mathcal{L}_t\sigma_{\phi}(t)^{1/2}) \|_{tr}.
\end{align}
Integrating over the time, we obtain the QSL limit
\begin{align}
  \tau\geq \tau_{QC}=\text{max}\{\frac{1}{\Lambda_{op}},\frac{1}{\Lambda_{tr}} \} 2\Delta\mathcal{Q} \left(1-\frac{2\mathcal{D}(\rho_0)\mp\Delta\mathcal{Q} }{2}\right)
\end{align}
where $\Lambda_{op,tr}^{\tau}=1/\tau\int_0^{\tau}dt(\|\mathcal{L}_t\rho_{\psi}(t)^{1/2} \|_{op,tr} +\|\mathcal{L}_t\sigma_{\phi}(t)^{1/2}) \|_{op,tr})$.
\subsection{Mandelstamm-Tamm bound:}
To derive MT bound, we rewrite the inequality Eq. (\ref{Boundeq}) as 
\begin{align}
2(1-\mathcal{D}(\rho_0)\mp\Delta\mathcal{Q})  \frac{d}{dt}\Delta\mathcal{Q} \leq \sqrt{\text{Tr}\{\mathcal{L}_t\rho_{\psi}(t)^{1/2}\mathcal{L}_t\rho_{\psi}(t)^{1/2 \dagger}\}} +\sqrt{\text{Tr}\mathcal{L}_t\sigma_{\phi}(t)^{1/2}\mathcal{L}_t\sigma_{\phi}(t)^{1/2 \dagger}}. \nonumber
\end{align}
Here, purity of the states $\rho_{\psi}(t)$ and  $\sigma_{\phi}(t)$ are unity. Using the definition of Hilbert-Schmidt norm of an operator, the MT bound as 
\begin{align}
2(1-\mathcal{D}(\rho_0)\mp\Delta\mathcal{Q})  \frac{d}{dt}\Delta\mathcal{Q} \leq \| \mathcal{L}_t\rho_{\psi}(t)^{1/2}\|_{HS}+\|\mathcal{L}_t\sigma_{\phi}(t)^{1/2}\|_{HS} \nonumber
\end{align}
and integrating over time, we obtain the bound as 
\begin{align}
  \tau\geq \tau_{QC}=\frac{1}{\Lambda_{HS}}2\Delta\mathcal{Q}\left(1-\frac{2\mathcal{D}(\rho_0)\mp\Delta\mathcal{Q}}{2} \right)
\end{align}
where $\Lambda_{HS}=1/\tau \int_0^{\tau}dt(\| \mathcal{L}_t\rho_{\psi}(t)^{1/2}\|_{HS}+\|\mathcal{L}_t\sigma_{\phi}(t)^{1/2}\|_{HS})$. 
Combining the above two bounds, we have 
\begin{align}
 \tau_{QC}=\text{max}\{\frac{1}{\Lambda_{op}},\frac{1}{\Lambda_{tr}},\frac{1}{\Lambda_{HS}} \} 2\Delta\mathcal{Q} \left(1-\frac{2\mathcal{D}(\rho_0)\mp\Delta\mathcal{Q} }{2}\right) 
\end{align}

\section{Dynamics under OU dephasing noise}
\label{sec4}
In what follows, we study the dynamics of newly-minted discord under OU dephasing noise and compare with the entanglement (quantified by concurrence) \cite{Con}. Further, we demonstrate the speed limit time bound on the decay and creation of quantum correlation in terms of affinity discord.  In general, the influence of environmental noise on the system is accounted by   completely positive trace-preserving operation namely called Kraus operator sum representation and can be written as \cite{Nielsen2010}
\begin{align}
  \rho(t)=\sum_{\mu}K_{\mu}(t)\rho(0)K_{\mu}^{\dagger}(t) 
\end{align} 
where $\rho(0)$ is the initial state, which is assumed to be free from the noise and the Kraus operators  $K_{\mu}$ satisfy $\sum_{\mu}K_{\mu}K_{\mu}^{\dagger}=\mathds{1}$. The Kraus operators of the Ornstein-Uhlenbeck  dephasing noise: $K_1=E_1^a\otimes E_1^b$, $K_2=E_1^a\otimes E_2^b$, $K_3=E_2^a\otimes E_1^b$ and $K_4=E_2^a\otimes E_2^b$ and the operators are
\begin{equation}
E_1^{a(b)}=
\begin{pmatrix}
p_{a(b)} & 0 \\
0 & 1 
\end{pmatrix},~~~ \text{and}~~~
E_2^{a(b)}=
\begin{pmatrix}
q_{a(b)} & 0 \\
0 & 0 
\end{pmatrix}. \nonumber
\end{equation}
where $q_{a(b)}=\sqrt{1-p^2_{a(b)}}$, $p_{a(b)}=\text{exp}(-f(t))$ and $f(t)=\frac{\Gamma t}{2}[1+(e^{-\gamma t}-1)/\gamma t]$. For our detailed analysis, we consider the Bell diagonal state 
\begin{align}
\rho(0)=\frac{1}{4} \left( \mathds{1}_a\otimes\mathds{1}_b+\sum_{j=1}^{3}c_{j}\sigma ^{j}_{a}\otimes\sigma ^{j}_{b} \right),
\label{case2}
\end{align}
where $-1\leq c_i=\langle \sigma_i \otimes\sigma_i \rangle \leq 1$ are the elements of $\vec{c}$ specifies the quantum state. It is worth pointing that the evolution of the under the OU noise preserve the structure of the initial state. The coefficients of the time evolved state are \cite{Evolve}
\begin{align}
  c_1(t)=c_1(0)e^{-2f(t)}, c_2(t)=c_2(0)e^{-2f(t)}, ~~~\text{and} ~~~c_3(t)=c_3(0). \nonumber
\end{align}
 The entanglement of the above mixed state is computed using the concurrence \cite{Con} and is defined as $C(\rho)=\text{max}\{0, \eta_1-\eta_2-\eta_3-\eta_4 \} $ with $\eta_i$ are the eigenvalues of the  matrix $\sqrt{\sqrt{\rho}\tilde{\rho} \sqrt{\rho}}$, $\tilde{\rho}=(\sigma_y\otimes\sigma_y )\rho^*(\sigma_y\otimes\sigma_y ) $ is spin flipped density matrix and * denotes the complex conjugation in the computational basis.  Now, we investigate the dynamics of affinity-based discord under OU dephasing noise and compare with the entanglement quantified by the concurrence. For this purpose, we consider the maximally entangled state as an initial state for our investigation.  
\begin{figure*}[!ht]
\centering\includegraphics[width=0.3\linewidth]{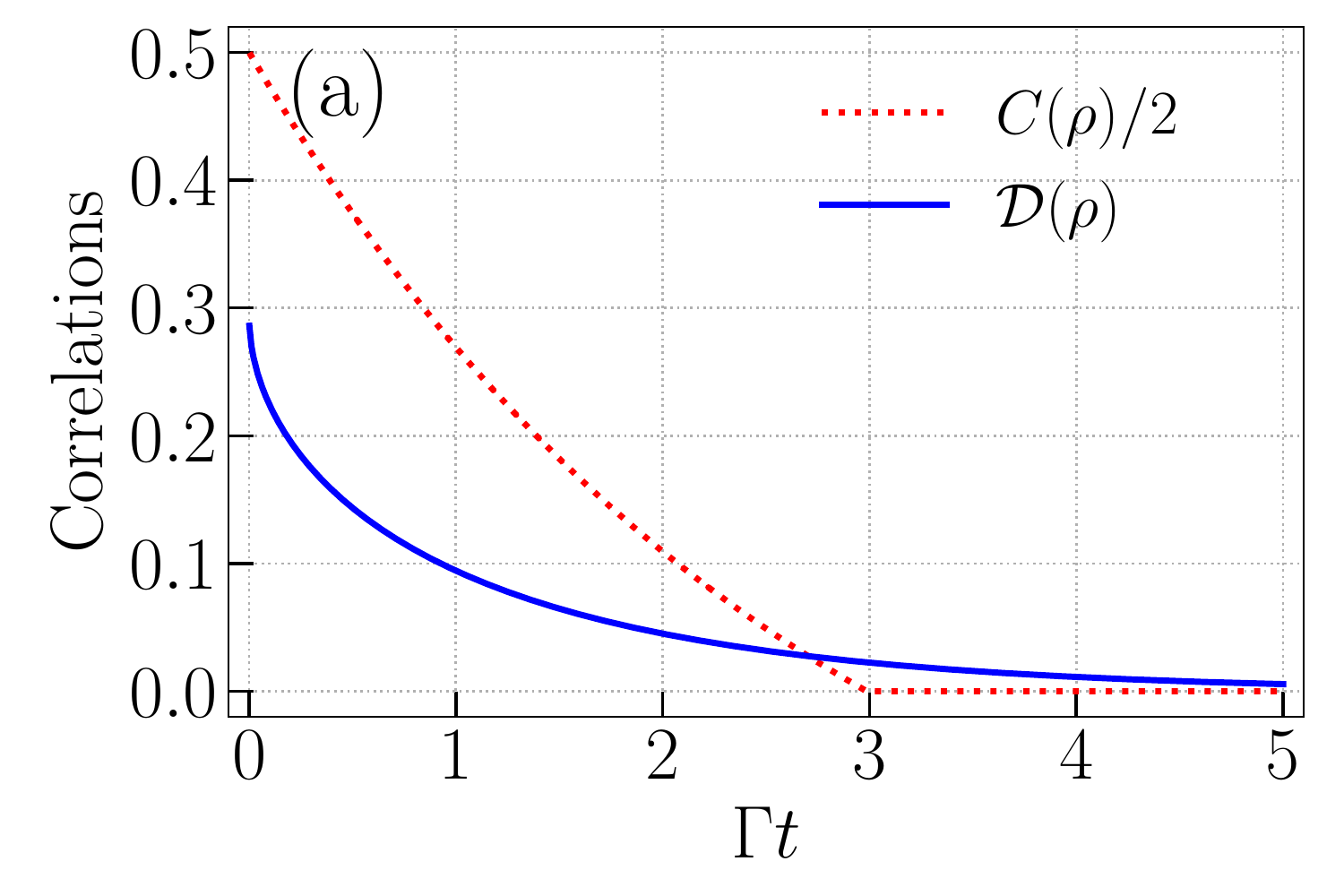}
\centering\includegraphics[width=0.3\linewidth]{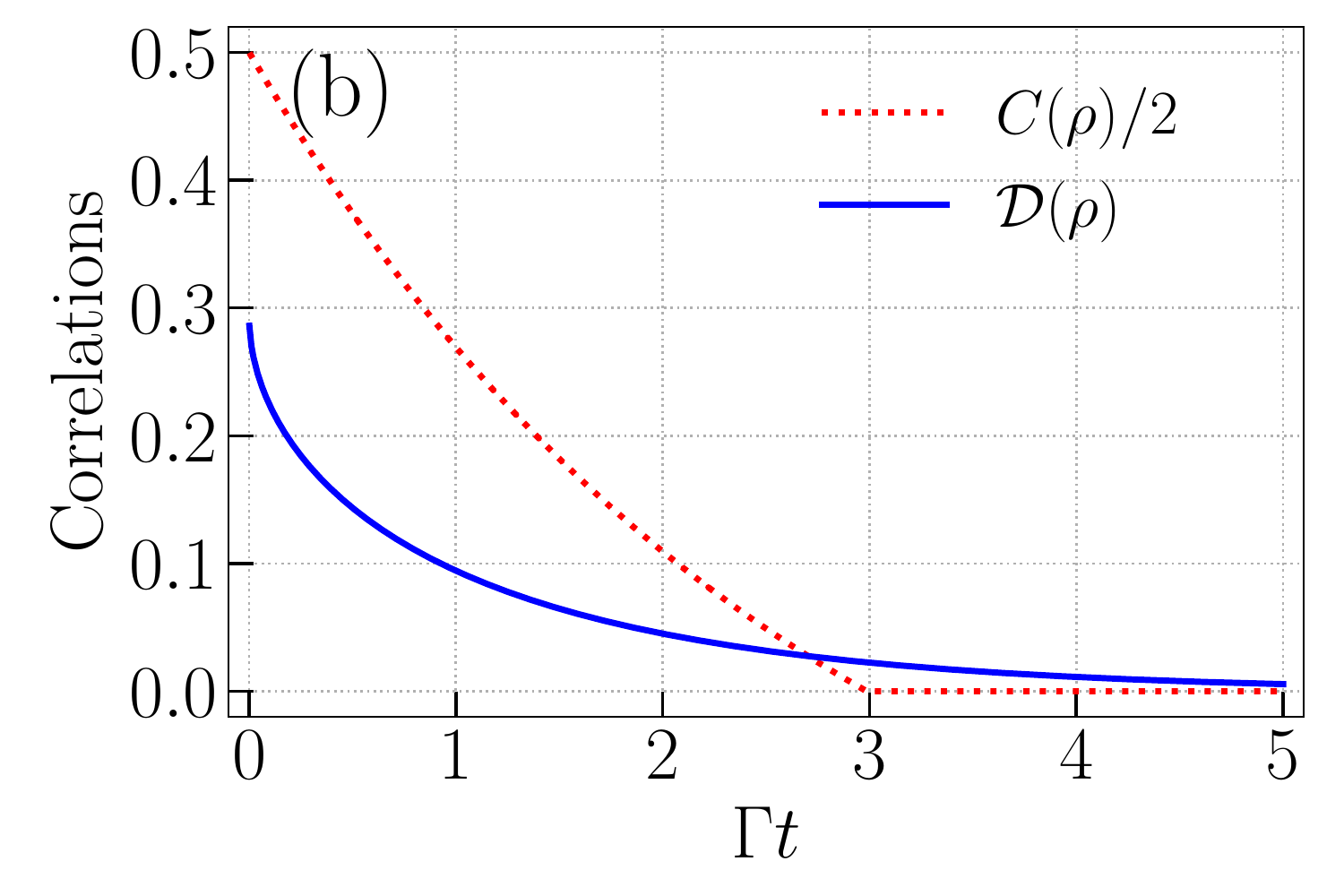}
\centering\includegraphics[width=0.3\linewidth]{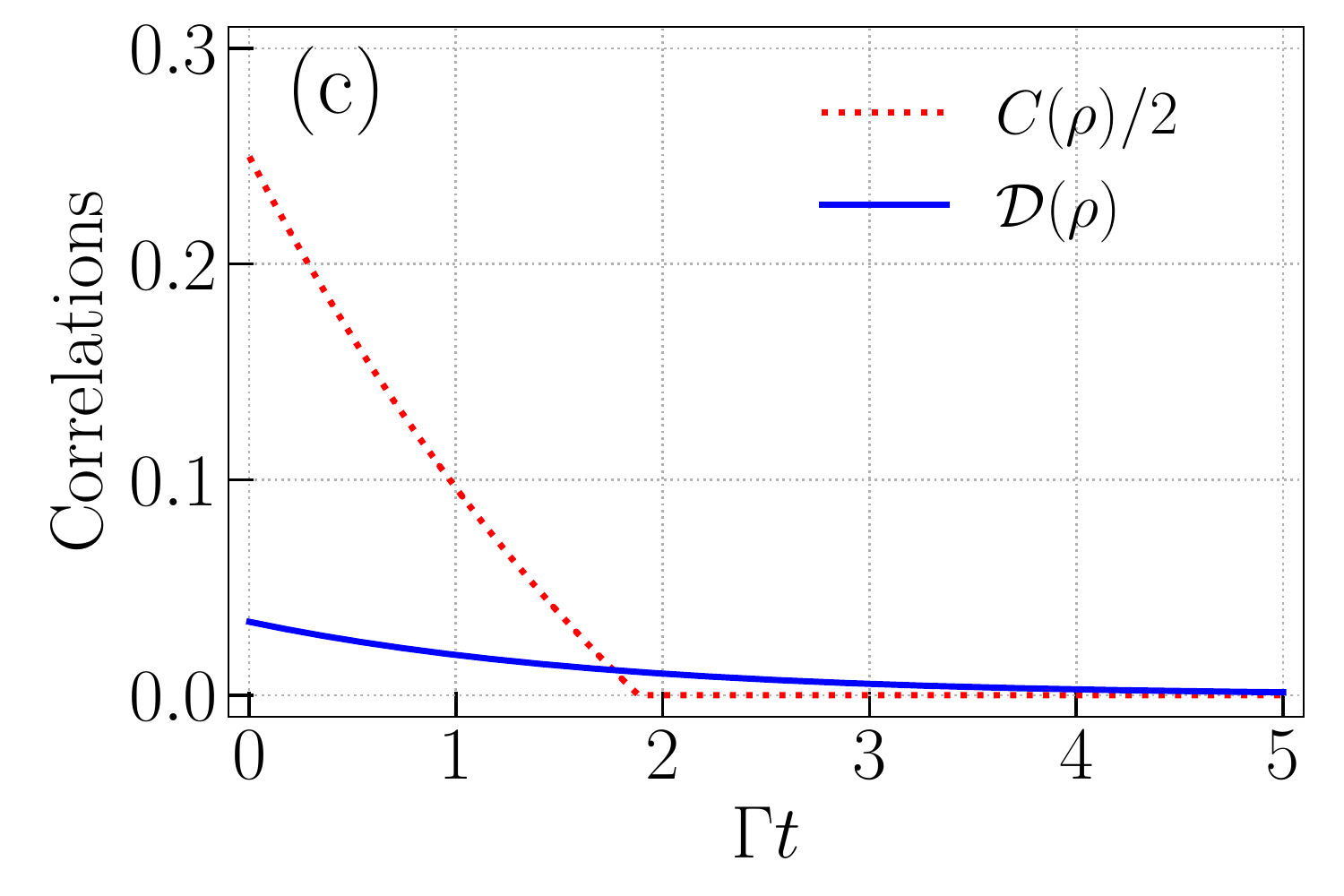}
\caption{(color online)  The dynamical behaviors of entanglement and affinity based measure under OU dephasing noise for the initial states (a)~ $(1,1,-1)$,~~ (b)~ $(1,-1,1)$, ~~ and ~~(c) ~ $(1,0.5,-0.5)$. The fixed parameter $\gamma=1$. }
\label{fig1}
\end{figure*}

We have plotted the dynamical behaviors of entanglement measured through concurrence and discord in Fig.1 for different initial conditions. In all the cases, we observe that the correlation measures decrease monotonically and the OU dephasing noise cause sudden death in entanglement whereas the affinity measure exhibit more robustness than the entanglement. The better robustness suggests that the affinity-based discord may provide advantages in information processing over the entanglement. 
\begin{figure*}[!ht]
\centering\includegraphics[width=0.5\linewidth]{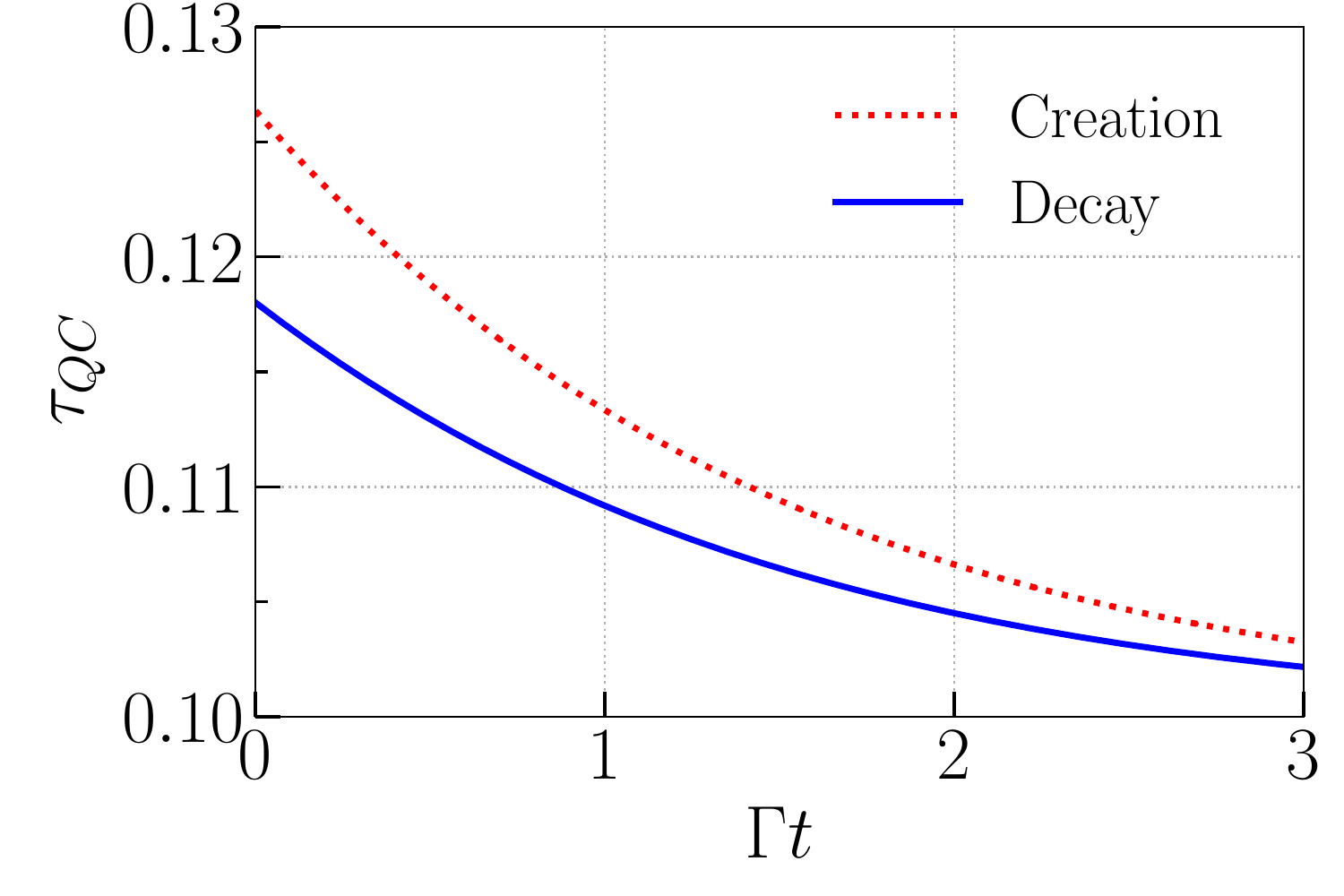}
\caption{(color online) QSL time for the creation and decay of quantum correlation under  OU dephasing noise.The fixed parameter $\gamma=1$.}
\label{QSLfig}
\end{figure*}

Considering the pure state with $(1,1,-1)$, we now study the QSL time limit of decay and creation of quantum correlation by an amount  $\Delta \mathcal{Q}$ through the affinity based measure. Here also we consider the above maximally entangled state for investigation. For pure state, the affinity is computed as $\mathcal{A}(\rho,\Pi^a(\rho))=\sum_ks_k^2$ with $s_k$ are the Schmidt coefficients and discord of maximally entangled state is $\mathcal{D}(\rho)=1-1/\sqrt{2}$.  In Fig. (\ref{QSLfig}), we have plotted the QSL time for the decay and  creation  of affinity based quantum correlations under the OU dephasing noise. Both decay and creation QSL decreases with the increase of coupling strength.

\section{Conclusions}
\label{cncl}
To summarize, we have defined a new variant of geometric version of quantum correlation measure using affinity and remedied the local ancilla problem of Hilbert-Schmidt based measures. We have derived the  Margolus-Levitin (ML) and Mandelstamm-Tamm (MT) bounds for the quantum speed limit time for the creation and decay of quantum correlation through the affinity measure. The better robustness of affinity based quantum  discord suggests that implementing some quantum information processing tasks with quantum correlation could be better than with entanglement under Ornstein-Uhlenbeck  noisy environment. 

Moreover, the influence of quantum correlation on quantum speed limit time is also studied under non-markovian evolution. It is worth-pointing that the affinity based metric used here also useful in quantification of coherence. Hence, the relation between the affinity based quantum coherence and  QSL time is worth investigation.

\section*{Acknowledgements}
Authors are indebted to the referees for their critical comments to improve the contents of the manuscript. RM is grateful for the financial support from MŠMT RVO 14000.



\end{document}